\documentclass[5p,times]{elsarticle}

\usepackage{amssymb,amsfonts,amsmath,graphicx,physics,hyperref}

\usepackage{ulem}

\journal{Journal of \LaTeX\ Templates}

\newcommand{\VEV}[1]{\langle #1 \rangle}
\newcommand{\bfrac}[2]{{\left(\frac{#1}{#2} \right)  }}

\newcommand\mpc{\,{\rm Mpc}}
\newcommand{\mpl}{m_{\rm Pl}}
\newcommand{\lfs}{\lambda_{\rm fs}}
\newcommand{\calL}{{\cal L}}
\newcommand{\calO}{{\cal O}}

\newcommand\gev{\,{\rm GeV}}
\newcommand\kev{\,{\rm keV}}
\newcommand\unitev{\,{\rm eV}}

\usepackage{eso-pic}
\newcommand{\reportnum}[2]{
  \AddToShipoutPictureBG*{%
    \AtPageUpperLeft{%
      \hspace{0.8\paperwidth}%
      \raisebox{#1\baselineskip}{%
        \makebox[0pt][l]{\textnormal{#2}}
  }}}%
}

\begin{document}

\reportnum{-4}{APCTP-Pre2023-035}
\reportnum{-5}{EPHOU-23-008}

\begin{frontmatter}

\title{Light cold dark matter from non-thermal decay}

\author[SKKU]{Ki-Young Choi}
\ead{kiyoungchoi@skku.edu}

\author[EWU,APCTP]{Jinn-Ouk Gong}
\ead{jgong@ewha.ac.kr}

\author[SKKU]{Junghoon Joh}
\ead{ttasiki1@skku.edu}

\author[JBNU,CSIC,VU]{Wan-Il Park}
\ead{wipark@jbnu.ac.kr}

\author[HU]{Osamu Seto}
\ead{seto@particle.sci.hokudai.ac.jp}

\address[SKKU]{Department of Physics and Institute of Basic Science, Sungkyunkwan University, Suwon 16419, Korea}
\address[EWU]{Department of Science Education, Ewha Womans University, Seoul 03760, Korea}
\address[APCTP]{Asia Pacific Center for Theoretical Physics, Pohang 37673, Korea}
\address[JBNU]{Division of Science Education and Institute of Fusion Science, Jeonbuk National University, Jeonju 54896, Korea}
\address[CSIC]{Instituto de Física Corpuscular, CSIC-Universitat de València, Paterna 46980, Spain}
\address[VU]{Departament de Física Teòrica, Universitat de València, Burjassot 46100, Spain}
\address[HU]{Department of Physics, Hokkaido University, Sapporo 060-0810, Japan}

\begin{abstract}
We investigate the mass range and the corresponding free-streaming length scale of dark matter produced non-thermally from decay of heavy objects which can be either dominant or sub-dominant at the moment of decay. We show that the resulting dark matter could be very light well below keV scale with a free-streaming length satisfying the Lyman-$\alpha$ constraints. We demonstrate two explicit examples for such light cold dark matter.
\end{abstract}


\end{frontmatter}

\section{Introduction}
\label{sec:intro}

The concordant cosmological model -- cold dark matter with cosmological constant, viz. $\Lambda$CDM, has been extremely successful in explaining the observed cosmic structure, from the cosmic microwave background~\cite{Planck:2018vyg} to galaxy scales~\cite{Chabanier:2019eai}. The cosmological requirements for dark matter (DM) are that its thermal velocity is negligible compared to the speed of light and that
its interactions to ordinary SM particles should be weak enough.
DM decouples from thermal plasma and develops gravitational potential well before the moment of the last scattering, accelerating the formation of hierarchical structure (see e.g.~\cite{Dodelson:2003ft}).

Since the thermal velocity of CDM particles is negligible, they can clump on scales even smaller than that of typical galaxies. This leads to copious sub-galactic haloes and dwarf galaxies, which has brought tensions with the observations on the abundance of small galaxies and dwarf satellites in the Local Group: The number of small bound objects is less than the predicted count of DM haloes~\cite{Bullock:2017xww}. Thus, to explain this deficiency of small-scale structure we need to suppress the amplitude of density perturbation below a certain cutoff scale.

Warm DM is an alternative to CDM to alleviate small-scale problems~\cite{Bode:2000gq}. Warm DM particles can travel a certain distance after its decoupling to the moment of the matter-radiation equality. This ``free-streaming length'' $\lfs$ provides a natural cutoff scale in the matter power spectrum, since perturbations on scales below $\lfs$ are erased as DM particles propagate without interruption. The most stringent constraint on $\lfs$ comes from the Lyman-$\alpha$ observations at redshift $2\lesssim z\lesssim5$, since the hydrogen Lyman-$\alpha$ radiation from intergalactic medium can trace density perturbations and underlying DM distribution that cannot be probed by any other means. Currently, the tightest lower bound from the Lyman-$\alpha$ observations on the thermal mass of the DM particles is around keV scale, which corresponds to $\lfs \sim \calO(1)$ Mpc~\cite{Irsic:2017ixq,DES:2020fxi,Villasenor:2022aiy}.

A well-known way of producing DM particles in the early universe is the freeze-out from thermal equilibrium~\cite{Lee:1977ua}. DM particles, once in thermal equilibrium with thermal plasma, become non-relativistic and decouple as the Universe expands. Many DM candidates are produced in this way, including weakly interacting massive particles (WIMPs). Even if the interaction of DM particles with thermal plasma is too weak to become thermalized, they still can be produced from thermal bath by scattering or decay of thermal particles. Such candidates include gravitino~\cite{Bolz:2000fu,Roszkowski:2004jd}, axino~\cite{Rajagopal:1990yx} and Dirac-type right-handed scalar neutrino~\cite{Asaka:2005cn}, also known as feebly interacting massive particles (FIMPs)~\cite{Baer:2014eja}. These ``thermally produced particles'' have naturally the same distribution in temperature and momentum as that of mother particles when they are produced. If such particles are heavier than keV, the corresponding $\lfs$ is smaller than Mpc scale and is consistent with the Lyman-$\alpha$ constraints.

However, DM particles produced non-thermally do not need to follow the constraints on the thermal mass. Then such particles are free from the Lyman-$\alpha$ mass constraints and can be very light far below keV scale, yet serve as CDM. We may call such DM as ``light cold dark matter (LCDM)''. Axion, or axion-like particles under bosonic coherent motion are one of the natural LCDM candidates even with very light mass down to $10^{-23}$ eV~\cite{Hui:2016ltb}. Recently, there have been some studies on how to obtain LCDM and what are the properties. For example, hidden axion from the parametric resonance of the associated scalar field, saxion~\cite{Co:2017mop, Harigaya:2019qnl}, and FIMP from freeze-in and Super WIMPs~\cite{Lopez-Honorez:2022sge} are LCDM candidates. In this article, we study systematically the general possibilities of LCDM produced from the decay of heavy non-thermal particles. We classify the production mechanism and the corresponding allowed bound of the DM mass $m_\chi$ or equivalently the free-streaming length scale $\lfs$.

This work is outlined as follows. In Section~\ref{sec:CDM}, we give the general discussions on the range of $m_\chi$ and $\lfs$ for both thermally and non-thermally produced particles. In Section~\ref{sec:NEP} we present more detailed analysis of non-thermally produced LCDM from the decay of heavy particles. We show explicit examples in Section~\ref{sec:Ex} of axion DM from saxion decay and DM from $Q$-ball decay. Then we conclude in Section~\ref{sec:con}.

\section{Mass and free-streaming length of dark matter}
\label{sec:CDM}

Regarding structure formation on small scales, our primary constraint will be the free-streaming length $\lfs$. In the standard big bang cosmology, it is given by~\cite{Kolb:1990vq,Borgani:1996ag}
\begin{equation}
\lfs
\equiv
\int_{t_{\rm dec}}^{t_{\rm eq}} \frac{v(t')}{a(t')} dt'  
\simeq 
\frac{t_{\rm NR}}{a_{\rm NR}} \left[ 2 + \log \bigg(\frac{t_{\rm eq}}{t_{\rm NR}} \bigg) \right] \, .
\label{eq:Lfs}
\end{equation}
Here, $t_{\rm dec}$ and $t_{\rm eq}$ denote respectively the decoupling time and the matter radiation equality, and the subscript ``NR'' denotes the epoch when DM becomes non-relativistic, which occurs when its momentum $p_\chi$ is similar to its mass, $p_\chi \sim m_\chi $. In the case of the DM particles that are produced by the decay of unstable particles, this formula is practically the same~\cite{Lin:2000qq}. From that time, the velocity is inversely proportional to the scale factor $a$. The overall magnitude of $\lfs$ is determined by the ratio $t_{\rm NR}/a_{\rm NR}$ by the epoch when DM becomes non-relativistic. The logarithmic correction that comes after $t_{\rm NR}$ is of the same order of magnitude [see \eqref{eq:logcorrection}].

To estimate $t_{\rm NR}/a_{\rm NR}$, we note that the time and the scale factor is related to the background temperature in the radiation-dominated era as
\begin{align}
t & = \frac{1}{2H}
\simeq 
1.32 \times 10^6 \sec \, \bfrac{3.36}{g_{*}(T)}^{1/2} \bfrac{\kev}{T}^2 \, ,
\\
\frac{a}{a_0}
& = \bfrac{g_{*s}(T_0)}{g_{*s}(T)}^{1/3}  \bfrac{T_0}{T}
\simeq 
2.35\times 10^{-7} \, \bfrac{3.91}{g_{*s}(T)}^{1/3} \bfrac{\kev}{T} \, ,
\end{align}
where $g_*(T)$ and $g_{*s}(T)$ are respectively the effective  degrees of freedom of the energy density and the entropy density at temperature $T$. Here we have used the present scale factor $a_0=1$,  the temperature $T_0=2.725  {\rm K}$, $g_*(T_0) = 3.36$ and $g_{*s}(T_0)=3.91$. Therefore, we can write down the free-streaming length \eqref{eq:Lfs} as
\begin{align}
\lfs
& \simeq 0.1 \mpc \, \bfrac{3.36}{g_{*}(T_{\rm NR})}^{1/2} \bfrac{3.91}{g_{*s}(T_{\rm NR})}^{-1/3} \bfrac{\kev}{T_{\rm NR}} 
\nonumber\\
&\quad \times \left[ 1 + \frac12 \log \bfrac{t_{\rm eq}}{t_{\rm NR}} \right] \, .
\label{eq:Lfs_TNR}
\end{align}
The log term for $t_{\rm NR}=10^6\sec$ is, with $t_{\rm eq}=51.1$ kyr,
\begin{equation}
\label{eq:logcorrection}
1+\frac12 \log\bfrac{t_{\rm eq}}{t_{\rm NR}} \simeq 8.15 \, .
\end{equation}

Now, we use the constraint from the Lyman-$\alpha$ observations on the free-streaming length in such a way that the length scale of Lyman-$\alpha$ $L_\alpha$ is greater than $\lfs$, with $L_\alpha \sim 0.1$ Mpc as a representative value. If the ratio $\lfs/L_\alpha$ is approximately of $\calO(1)$, DM is warm. If it is much bigger (smaller) than 1, it is hot (cold), respectively. If we require that $\lfs/L_\alpha \lesssim 1$, then Eq.~\eqref{eq:Lfs_TNR} translates into the following bound on $T_{\rm NR}$:
\begin{equation}
T_{\rm NR} \gtrsim 8.15 \kev \, \bfrac{0.1\mpc}{L_\alpha} \, ,
\label{TNRbound}
\end{equation}
with, for $T_{\rm NR}< 1$ MeV, $g_*(T_{\rm NR})=3.36$ and $g_{*s}(T_{\rm NR})=3.91$.

Up to now, our discussions on $\lfs$ are general. We now specify to thermally produced particles $\chi$ such as WIMPs. The ratio of the average momentum and the temperature of relativistic DM is $\VEV{p_\chi}/T_\chi \simeq 3.151$. After the decoupling of $\chi$ from thermal background, both $\VEV{p_\chi}$ and $T_\chi$ scale proportional to $1/a$, and thus the ratio $\VEV{p_\chi}/T_\chi \simeq 3.151$ does not change. On the other hand, the background temperature $T$ increases due to the annihilation of massive particles, and the ratio of the DM temperature $T_\chi$ to $T$ is given by
\begin{equation}
\frac{T_\chi}{T} = \bfrac{g_{*s}(T) }{g_{*s}(T_{\rm dec})}^{1/3} \, ,
\label{eq:TXT}
\end{equation}
assuming that the decoupled particles are relativistic and the total entropy is conserved. Finally, thermal DM becomes non-relativistic when $\VEV{p_\chi}\simeq m_\chi$, or $T_\chi \simeq m_\chi/3.151$. From Eq.~\eqref{eq:Lfs_TNR}, we obtain
\begin{align}
\lambda_{fs} 
& \simeq 
0.32  \mpc \, \bfrac{3.36}{g_{*}(T_{\rm NR})}^{1/2} \bfrac{3.91}{g_{*s}(T_{\rm NR})}^{-1/3}   
\bfrac{\kev}{m_\chi}\bfrac{T_\chi}{T}_{\rm NR}
\nonumber\\
&\quad \times \left[ 1 + \frac12 \log \bfrac{t_{\rm eq}}{t_{\rm NR}} \right] \, ,
\label{eq:Lfs_TPP}
\end{align}
where $T_\chi/T$ is given by Eq.~\eqref{eq:TXT}. For particles with mass around keV which decoupled, such as warm DM\footnote{Actually, in this case, the relic density is determined as~\cite{Viel:2005qj}
\begin{equation}
\Omega_\chi h^2 = \bfrac{10.75}{g_{*s}(T_{\rm dec})} \bfrac{m_\chi}{94 \unitev}.
\end{equation}
Therefore, the constraint from Lyman-$\alpha$ is incompatible with the relic density of DM unless $g_{*s}(T_{\rm dec}) \gtrsim 10^3$.
}, it can be cold ($\lfs \lesssim L_\alpha$) if its mass is as large as
\begin{equation}
m_\chi \gtrsim 8.66 \kev \, \bfrac{0.1\mpc}{L_\alpha}\bfrac{106.75 }{g_{*s}(T_{\rm dec})}^{1/3} \, .
\label{eq:WDM}
\end{equation}
This is applied to all thermal DM, which has a similar form of the distribution function as that of thermal particles. The only way to avoid the constraint on the mass of DM from Lyman-$\alpha$ in Eq.~\eqref{eq:WDM} is to consider non-thermal production of DM \footnote{If there is energy injection to the visible particles only after the decoupling of dark matter from thermal bath, it is possible to obtain sub-{\rm keV} cold dark matter.}.

Now, we move to the production of LCDM from the decay of non-thermal particles. We assume that DM is very weakly interacting so that it is not in thermal equilibrium when produced from the decay of heavier non-relativistic particle with the mass $M$. At the moment of production, they are relativistic with the momentum of the order of $M$~\footnote{For degenerate case where the mass difference of mother and daughter particles is very small, the daughter particle can be still non-relativistic too~\cite{Heeck:2017xbu,Boulebnane:2017fxw}.}, e.g. $p_\chi\simeq M/2$ for two-body decay. After then, $p_\chi$ redshifts due to the expansion of the Universe as
\begin{equation}
p_\chi = \frac{M}{2} \bfrac{a_{\rm prod}}{a}  
= \frac{M}{2} \bfrac{T}{T_{\rm prod}} \bfrac{g_{*s}(T) }{g_{*s}(T_{\rm prod})}^{1/3} \, ,
\end{equation}
where the subscript ``prod'' means the time of production. When $p_\chi=m_\chi$, DM becomes non-relativistic and the background temperature $T_{\rm NR}$ at the moment is given as
\begin{equation}
T_{\rm NR} = T_{\rm prod}\bfrac{2m_\chi}{M}\bfrac{g_{*s}(T_{\rm NR}) }{g_{*s}(T_{\rm prod})}^{-1/3} \, .
\label{eq:TNR_decay}
\end{equation}
By using Eq.~\eqref{eq:Lfs_TNR}, we obtain
\begin{align}
\lfs
& \simeq 0.03 \mpc \, \bfrac{3.36}{g_{*}(T_{\rm NR})}^{1/2} \bfrac{3.91}{g_{*s}(T_{\rm NR})}^{-2/3} 
\bfrac{g_{*}(T_{\rm prod})}{106.75}^{-1/3}  
\nonumber\\
& \quad 
\times \bfrac{\kev}{m_\chi} \frac{M}{2T_{\rm prod}} \left[ 1 + \frac12 \log \bfrac{t_{\rm eq}}{t_{\rm NR}} \right] \, .
\label{eq:Lfs_decay}
\end{align}
For example, if DM is produced when $T\gg 10^3 \gev$, and becomes non-relativistic after the big bang nucleosynthesis epoch, then the free-streaming length is given by
\begin{equation}
\lfs \simeq 0.27 \mpc \, \bfrac{\kev}{m_\chi} \frac{M}{2T_{\rm prod}} \, .
\label{eq:Lfs_high}
\end{equation}
In this case, DM can be cold ($\lfs \lesssim L_\alpha$) for the mass 
\begin{equation}
m_\chi \gtrsim 1.35 \kev \, \bfrac{M}{T_{\rm prod}}\bfrac{0.1\mpc}{L_{\alpha}} \, .
\label{eq:Lowmass}
\end{equation}
Therefore, if the mass of the decaying non-relativistic particle is much smaller than the background temperature at the time of the DM production, then DM with mass smaller than keV can be still cold enough, as noticed in Ref.~\cite{Moroi:2020has}.

%%%%%%%%%%%%%%%%%%%%%%%%%%%%%%%%%
\section{Dominance of mother particle and dark matter}
\label{sec:NEP}
%%%%%%%%%%%%%%%%%%%%%%%%%%%%%%%%%

Given general discussions on the free-streaming length $\lfs$ and the mass of DM $m_\chi$ from a decaying non-thermal heavy particle, here we consider in detail the mass range of LCDM from decay for two cases: Whether or not the decaying heavy particle is dominant at the moment of decay.

\subsection{Dark matter from decay of dominating heavy particle}

In this section, we consider LCDM produced from heavy non-relativistic particles which dominated the energy density of the Universe when they decay. Although LCDM from the inflaton decay has been considered in Ref.~\cite{Moroi:2020has}, the mother particle can be more general such as curvaton~\cite{Enqvist:2001zp,Lyth:2001nq,Moroi:2001ct}. Nevertheless, as a representative example, we use the term ``inflaton'' for the heavy decaying field.

The perturbative decay of the inflaton field is completed when the decay rate $\Gamma$ is similar to the Hubble expansion rate $H$, 
$\Gamma \sim H$, with 
\begin{equation}
H =  \frac{T_R^2}{\mpl}\sqrt{\frac{\pi^2}{90} g_*(T_R)} \, ,
\end{equation}
where $\mpl\equiv(8\pi G)^{-1/2}$ is the reduced Planck mass and $T_R$ is the reheating temperature after inflation given by  
\begin{equation}
T_R = \bfrac{90}{\pi^2 g_*(T_R)}^{1/4 } \sqrt{\Gamma\mpl} \, ,
\end{equation}
with the sudden decay approximation. $T_R$ is bounded from the tensor-to-scalar ratio of the Planck observation as~\cite{Planck:2018vyg} 
\begin{equation}
T_R < 5\times 10^{15}\gev \, .
\end{equation}
While most of the inflaton energy is transferred to the visible particles with the common temperature $T_R$, we assume that a small amount of energy goes into DM with a partial decay rate $\Gamma_\chi$, and that DM is so weakly interacting with visible sector that it is always decoupled from the time of production until now.

Assuming that most of DM production happens at around the last epoch of reheating, we can estimate $T_{\rm prod}\simeq T_R$, so that $M/T_{\rm prod}$ in Eq.~\eqref{eq:Lfs_high} is
\begin{equation}
\frac{M}{T_{\rm prod}} \simeq \frac{M}{T_{R}} 
= \bfrac{90}{\pi^2 g_*(T_{R})}^{-1/4 } \frac{M}{ \sqrt{\Gamma\mpl}} \, .
\end{equation}
Therefore, DM lighter than keV can be cold if $M\ll\sqrt{\Gamma\mpl}$ from Eq.~\eqref{eq:Lowmass}. If produced at an earlier stage of the reheating processes, DM can be colder because the background temperature can be higher than $T_R$. In addition, as was mentioned in Ref.~\cite{Moroi:2020has}, if there occurs amplification in the production of scalars due to the Bose-Einstein enhancement, then more particles can be produced in the earlier epochs.

There are other constraints from the relic density of DM and possible thermalization. If DM self-interactions are ignored, the non-thermal contribution from the decay of the zero-mode inflaton in the sudden decay approximation is given by
\begin{equation}
Y_\chi^\text{non-th} \equiv \frac{n_\chi}{s} \simeq \frac{3\alpha_\chi B_\chi T_R}{2M},
\label{eq:Ynon}
\end{equation}
where $s=2\pi^2g_{*s}T^3/45$ is the entropy density, $B_\chi\equiv \Gamma_{\phi\rightarrow \chi\chi}/\Gamma$ is the branching fraction and $\alpha_\chi \gtrsim 1$ is the possible enhancement factor from the amplification of the production due to the Bose-Einstein enhancement~\cite{Moroi:2020has}. Then, the relic density of DM is
\begin{equation}
\label{Oh2}
\Omega_{\chi} h^2 = 0.1 \bfrac{m_\chi Y_\chi^\text{non-th} }{3.7\times 10^{-11} \gev}  \simeq  0.1 \bfrac{0.27\mpc}{\lfs}\bfrac{\alpha_\chi B_\chi}{5\times 10^{-5}}\, ,
\end{equation}
where we used  Eq.~(\ref{eq:Lfs_high}) and $T_{\rm prod}=T_R$ in the last equation.

If $T_R > M$, the inflaton can be thermally produced by its pair creation and the inverse decay from thermal particles. Then, the processes mediated by the inflaton could thermalize DM particles $\chi$, which conflicts with our assumption mentioned above.
While the thermalization of $\chi$ via the thermal inflaton decay can be avoided, if $ \Gamma_\chi < H$, or 
\begin{equation}
\Gamma_{\phi \to \chi\chi} \ll \calO(10) \bfrac{M}{\mpl} M ,
\label{eq:G-chi-cond}
\end{equation}
is satisfied, the freeze-in production of $\chi$ can be still sizable. 
The expected abundance of thermal DM is given by~\cite{Choi:1999xm,Kim:2008yu}
\begin{equation}
\left. Y_\chi^{\rm th} \right|_{T = M/3} \simeq \frac{2.2}{g^{3/2}_*(M/3)} 
\frac{\Gamma_{\phi \to \chi \chi}\mpl}{M^2} \, .
\end{equation}
By comparing this with Eq.~\eqref{eq:Ynon}, we find that in order for the thermal contribution to be of only a small fraction $f_\chi^{\rm th} \ll 1$, $T_R$ is constrained as
\begin{equation}
T_R \lesssim \frac{2\sqrt{10} \alpha_\chi f_\chi^{\rm th} g_*^{3/2}(M/3)}{\pi g_*^{1/2}(T_{R})} M 
\simeq 20\alpha_\chi M \left(\frac{f_\chi^{\rm th}}{0.1}\right).
\label{eq:TR-cond}
\end{equation}
%

%%%%%%%%%%%%%%%%%%%   
\begin{figure*}[!t]
\begin{center}
\begin{tabular}{ccc} 
 \includegraphics[width=0.3\textwidth]{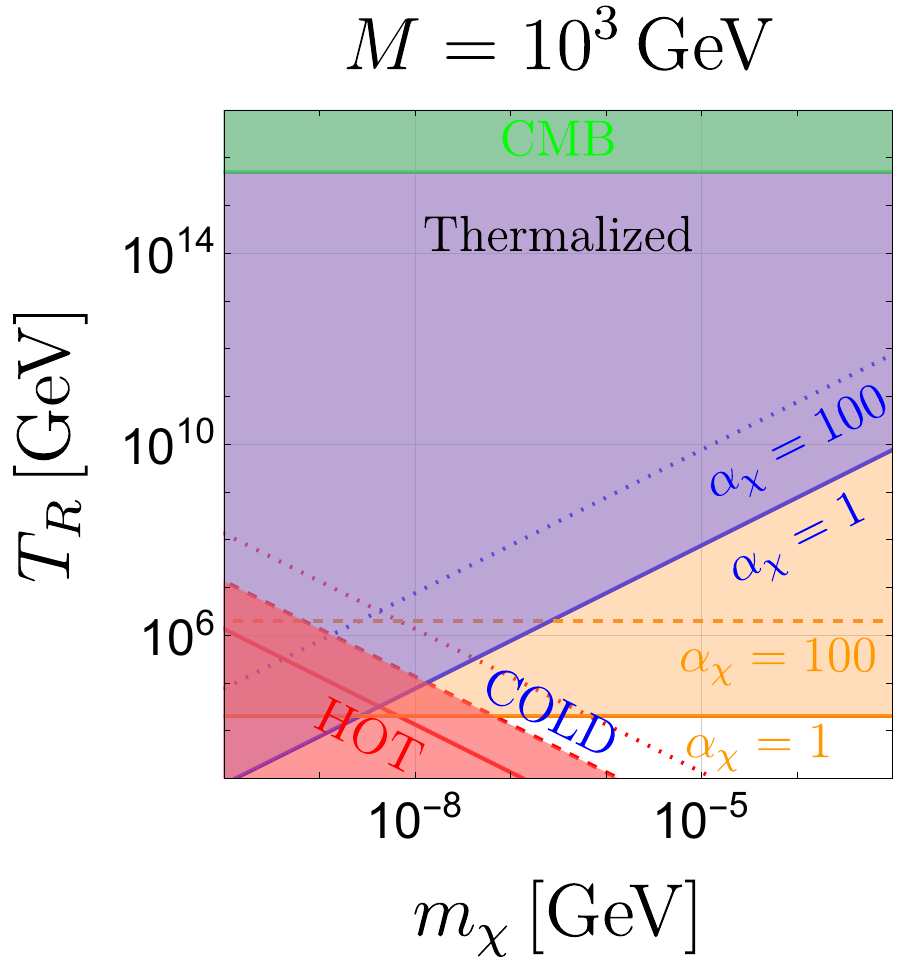}
 &
 \includegraphics[width=0.3\textwidth]{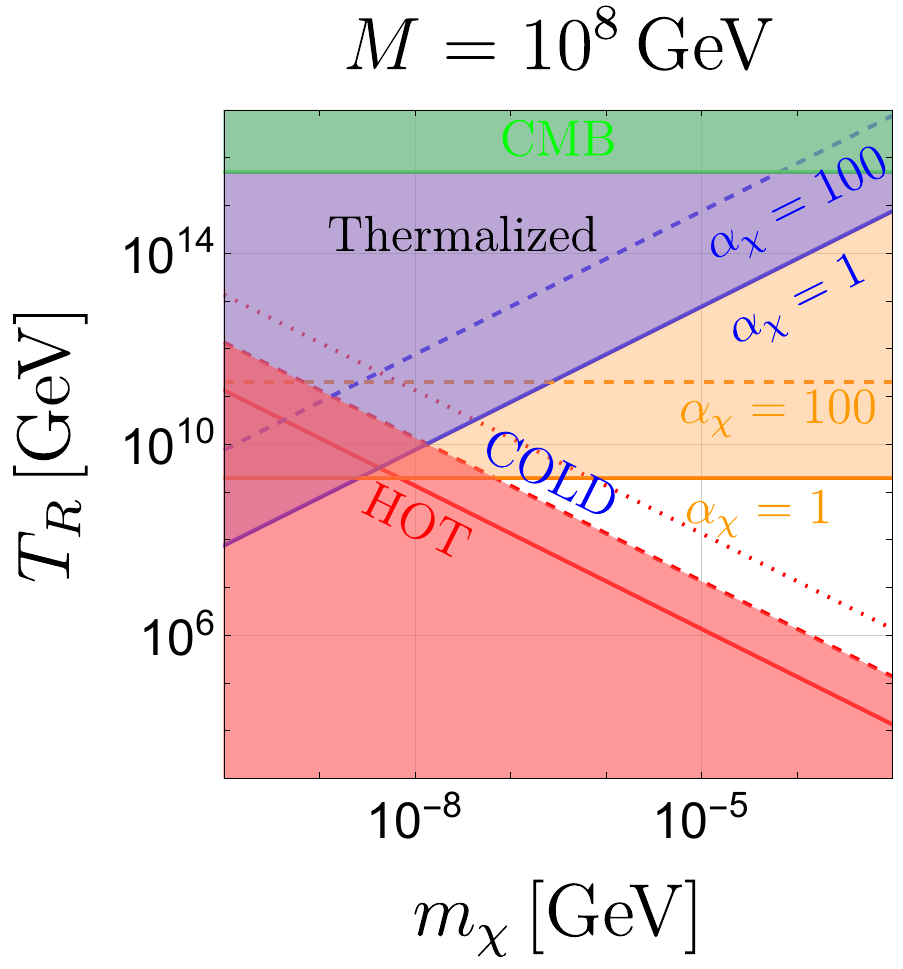}
 &
 \includegraphics[width=0.3\textwidth]{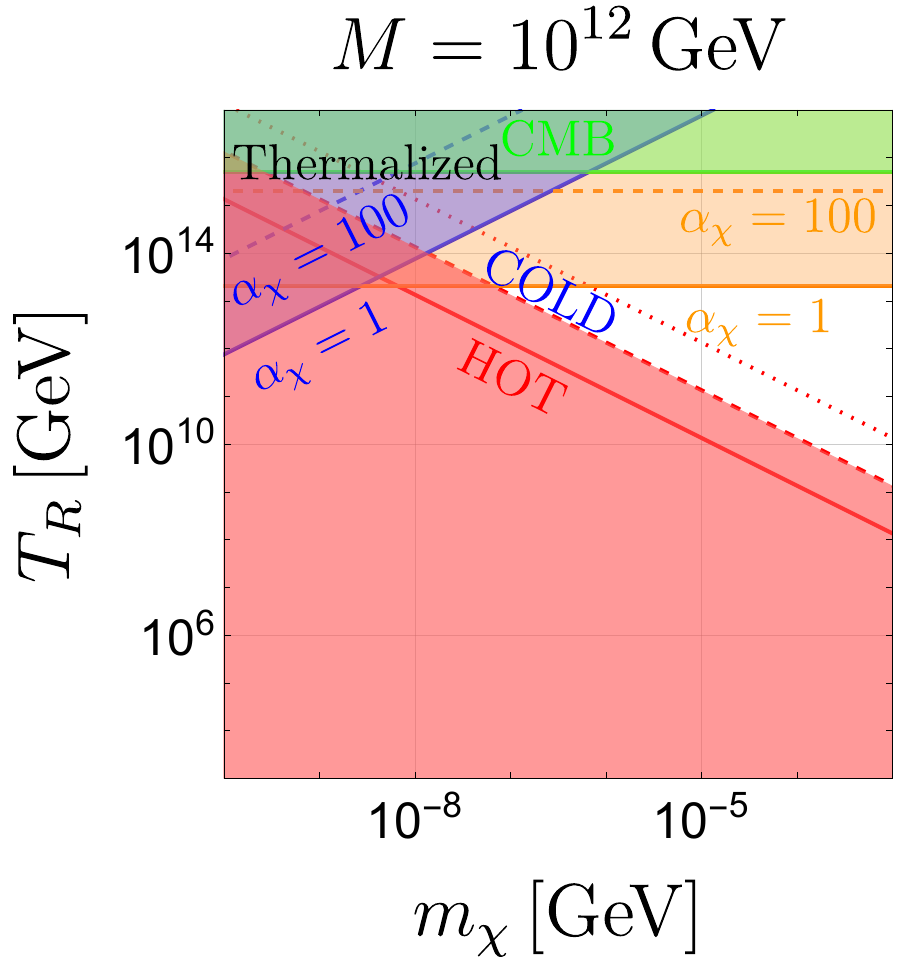}
  \end{tabular}
\end{center}
\caption{The free-streaming length of DM produced from the decay of inflaton with mass $M=(10^3,10^8,10^{12}) \gev$ (left, middle, right), respectively. Red lines denote $\lambda_{fs}=1, 0.1, 0.01\mpc$  (solid, dashed, and dotted, respectively) and become smaller for heavier mass of DM. Green line is the upper bound on $T_R$ from CMB, and blue line shows the bound from thermal equilibrium of DM. The yellow region is disfavored since the thermal DM to be larger than non-thermal ones. }
\label{fig:Lfs_dom}
\end{figure*}
%%%%%%%%%%%%%%%%%%%%%

Combining this with the DM relic density constraint \eqref{eq:Lowmass}, we find the lower bound on the DM mass from the inflaton decay as
\begin{equation}
m_\chi \gtrsim 68\unitev \, \bfrac{1}{\alpha_\chi}\bfrac{0.1
\mpc}{L_\alpha} \, ,
\label{eq:mchi_low1}
\end{equation}
 where we have used $f_\chi^{\rm th}=0.1$ as a reference value.
It is independent from the inflaton mass or the reheating temperature. However, when $\alpha_\chi $ is large, the constraint from Eq.~\eqref{eq:G-chi-cond} is stronger and the lower bound on the mass is 
\begin{equation}
m_\chi \gtrsim 13\unitev \, \bfrac{1}{\alpha_\chi}^{1/2}\bfrac{0.1
\mpc}{L_\alpha}^{1/2} \, .
\end{equation}
In Fig.~\ref{fig:Lfs_dom}, we show the constraints and the allowed region for LCDM in the plane of $(m_\chi,T_R)$.
Here, we assume that the non-thermal DM comprise 100\% of DM, which can be obtained by varying the parameters of $\alpha_\chi B_\chi$ as in Eq.~(\ref{Oh2}).

\subsection{Dark matter from decay of subdominant heavy particle}

Next, we consider the case that the decoupled heavy particle is subdominant when it decays. The background temperature at this time is given by
\begin{equation}
T_{\rm decay} \simeq \bfrac{\pi^2 g_*}{90}^{-1/4}\sqrt{\Gamma\mpl} \, .
\label{eq:Tdecay}
\end{equation}
We define $R_\phi \equiv \rho_\phi/\rho|_\text{decay}$ as the fraction of its energy density at this moment. Being subdominant, the condition $R_\phi\ll1$ holds. DM is produced from the decay of $\phi$ with the momentum of the order of the mass of the mother particle, e.g. $p_\chi=M/2$ for two-body decay.

As discussed in the previous section, $\lfs$ is given by Eq.~\eqref{eq:Lfs_decay}, with $T_{\rm prod}=T_{\rm decay}$ given in Eq.~\eqref{eq:Tdecay}. However, the difference from the previous section is that now the energy density of $\phi$ is subdominant, and the DM relic density \eqref{Oh2} is modified to
\begin{equation}
\label{Oh2_hidden}
\Omega_{\chi} h^2 \simeq 0.1 \bfrac{0.27\mpc}{\lfs} \bfrac{\alpha_\chi B_\chi R_\phi}{5\times 10^{-5}}\, .
\end{equation}

{\bf Visible decay}\\
If the heavy scalar decays dominantly to the visible particles so that $B_\chi \ll 1$, and $M$ is smaller than the background temperature, it would be easily thermalized with the thermal plasma since $\Gamma > H$. To avoid the rethermalization of DM with the thermal $\phi$, the decay rate of $\phi$ into DM must be smaller than the expansion rate, $\Gamma_{\phi \rightarrow \chi\chi} < H$. Similar to the case of dominant decay, the lower bound on the DM mass as
\begin{equation}
m_\chi \gtrsim 68 \unitev \, \bfrac{1}{\alpha_\chi R_\phi}\bfrac{0.1 \mpc}{L_\alpha} \, .
\label{eq:mchi_low2}
\end{equation}
Since $R_\phi <1$, the lower bound on the DM mass becomes larger than the dominant decay of $\phi$ in Eq.~\eqref{eq:mchi_low1}. \\

{\bf Hidden decay}\\
If the scalar decays dominantly to DM in the hidden sector ($B_\chi\simeq1$), the interaction of $\phi$ with the visible matter might be very small and $\phi$ may not be thermalized with the visible sector. In this case the inverse decay discussed in the previous section can be ignored.

\begin{figure*}[!t]
\begin{center}
\begin{tabular}{ccc} 
 \includegraphics[width=0.3\textwidth]{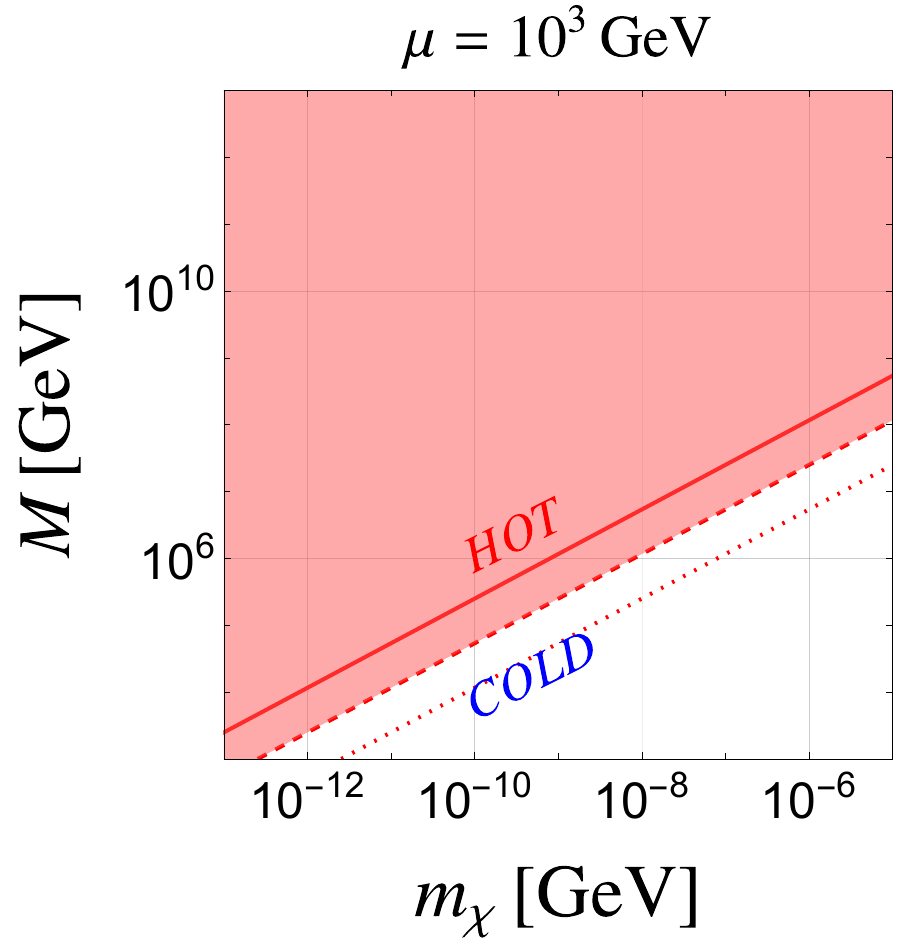}
 &
 \includegraphics[width=0.3\textwidth]{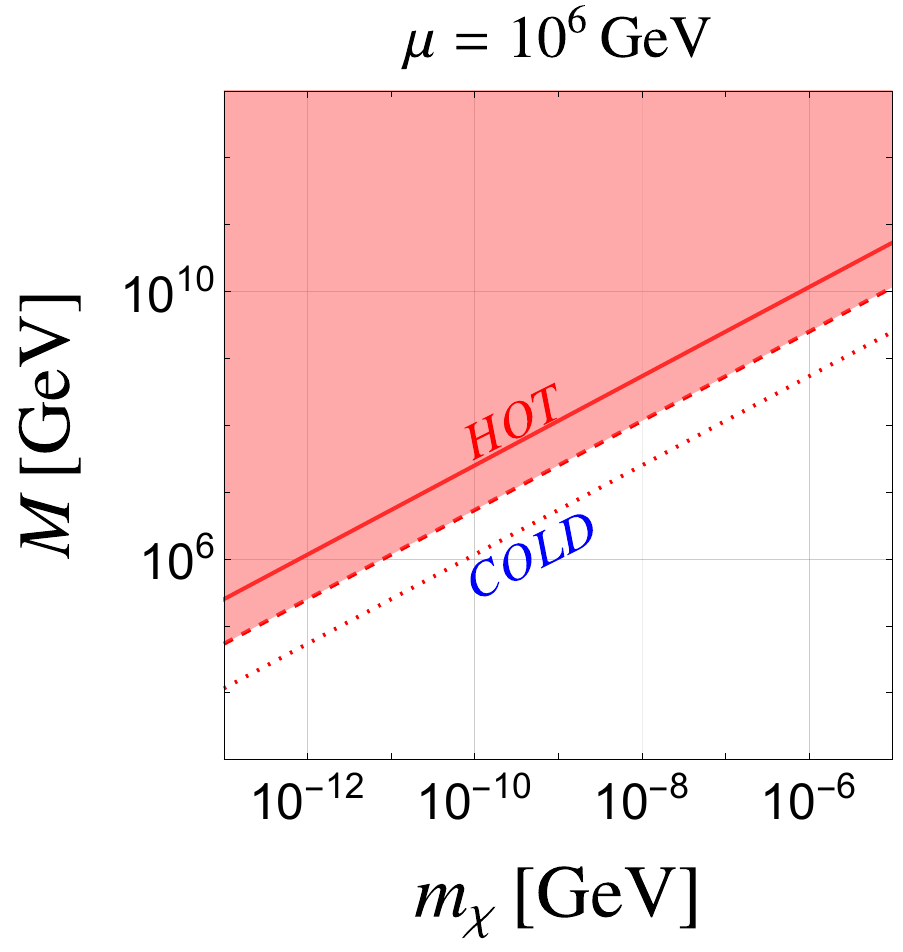}
 &
 \includegraphics[width=0.3\textwidth]{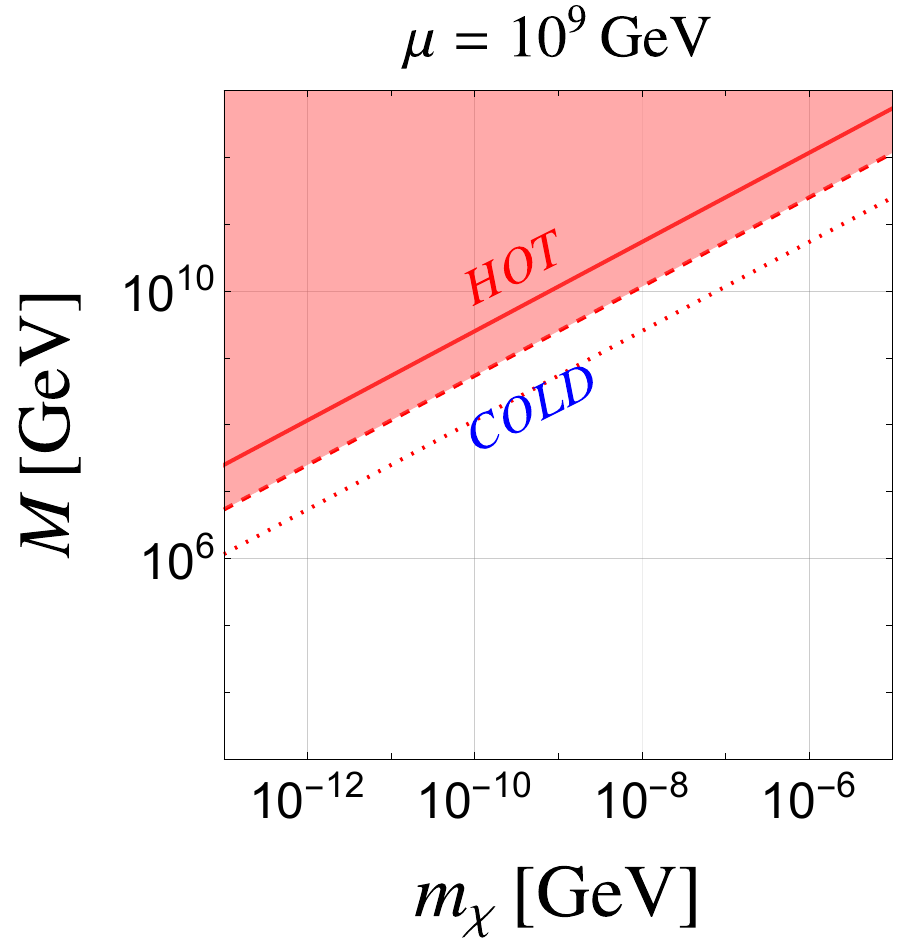}
  \end{tabular}
\end{center}
\caption{The free-streaming length of DM produced form the decay of subdominant heavy particle in the plane of DM mass $m_\chi$ and the heavy particle mass $M$. Red lines denote $\lambda_{fs}=1, 0.1, 0.01\mpc$  (solid, dashed, and dotted, respectively). 
}
\label{fig:Lfs_subdom}
\end{figure*}

For the interaction Lagrangian $\calL_{\rm int} = \mu \phi\chi^2+g \phi h^2$  with scalar DM $\chi$ and the SM sector scalar $h$, the total decay rate is given by $\Gamma \simeq \Gamma_\phi = \mu^2/(8\pi M)$ for the hidden decay to DM,
if $ g^2\ll \mu^2$.   Then,  the free-streaming length is estimated as
\begin{equation}
\lfs \simeq 8 \times 10^{-2} \mpc \, \bfrac{\kev}{m_\chi} \, \bfrac{M}{10^{12}\gev}^{3/2} \, \bfrac{10^{10}\gev}{\mu} \, .
\end{equation}
It means that DM with mass lighter than keV can be cold for small $M$ and large $\mu$. In Fig.~\ref{fig:Lfs_subdom}, we show the constraints and the allowed region for LCDM in the plane of $(m_\chi,M)$, assuming that the whole DM is non-thermal component. For given $\lfs$, this can be obtained for appropriate values of $\alpha_\chi B_\chi R_\phi$. 

For this small mass of DM, the constraint from $\Delta N_{\rm eff} \equiv N_{\rm eff} - N_{\rm eff}^{\rm SM}$, the additional effective number of neutrino species compared to the SM value  $N_{\rm eff}^{\rm SM}=3.045$~\cite{deSalas:2016ztq}, becomes important for thermally produced hot DM. However, even though the non-thermal DM in our study might be relativistic around the epoch of BBN, its contribution to the energy density of the Universe is still quite small during the epoch.
Specifically, in our case, if $T_{\rm NR}$ is smaller than 1 MeV, then $\Delta N_{\rm eff} $ at the BBN epoch is estimated as
\begin{align}
\Delta N_{\text{eff}}
&= \frac{\rho_\chi}{\rho_\nu} 
 \simeq 5.40 \times 10^{-4} \bfrac{T_\text{eq}}{1\unitev} \bfrac{8.15\kev}{T_\text{NR}},
\end{align}
where we used that the energy density of DM and radiation is the same at matter-radiation equality temperature $T_{\rm eq}$. Considering the lower bound on $T_{\rm NR}$ from Eq.~(\ref{TNRbound}), $\Delta N_{\rm eff} $ is irrelevant compared to  the observational bound  given by $N_{\text{eff}}=2.88 \pm 0.27$~\cite{Pitrou:2018cgg}.

%%%%%%%%%%%%%%%%%%%   
%%%%%%%%%%%%%%%%%%%%%

\section{Examples of LCDM}
\label{sec:Ex}

In this section, we show examples of LCDM: Axion-like particle as DM from saxion decay and scalar DM produced from the decay of $Q$-ball.

\subsection{Hidden axion DM from saxion decay}

In supersymmetric axion model with spontaneously broken Peccei-Quinn (PQ) symmetry, the radial part of the scalar field is called the saxion $s$, and the phase component is the axion $a$. We consider a general hidden sector axion, which is different from the QCD axion. If this axion is stable until today, it can be a good candidate for DM. The mass of the axion may come from the PQ-symmetry breaking part in the hidden sector.

Here, we examine the possible mass of axion DM when it is produced from the decay of saxion in the early Universe. The saxion is initially displaced from the vacuum with $\delta s \simeq f_a$ with $f_a$ being the PQ scale, and begins oscillation when the Hubble expansion rate is similar to the saxion mass, $H\sim m_s$. Here, we simply assume that the potential energy of saxion is  $V= m_s^2 s^2/2$.
The saxion has the following interaction with the axion~\cite{Chun:1995hc}:
\begin{equation}
\left( 1+ \frac{\sqrt2 x}{f_a} s \right)  \frac12 \partial_\mu a \partial^\mu a \, ,
\end{equation}
where $x=\calO(1)$. The saxion may dominantly decay into two axions with a decay rate 
\begin{equation}
\Gamma_{s\rightarrow aa} \simeq \frac{x^2}{64\pi}\frac{m_s^3}{f_a^2} \, .
\end{equation}
The free-streaming length of the axion produced from the saxion decay is then given by
\begin{equation}
\lfs \simeq 7\times 10^{-2}\mpc \, \bfrac{\kev}{m_a}\bfrac{10^9\gev}{m_s}^{1/2} \bfrac{f_a}{10^{12}\gev} \, ,
\end{equation}
where we have set $x=1$.

%%%%%%%%%%%%%%%%%%%   
\begin{figure}[!t]
\begin{center}
\begin{tabular}{c} 
 \includegraphics[width=0.4\textwidth]{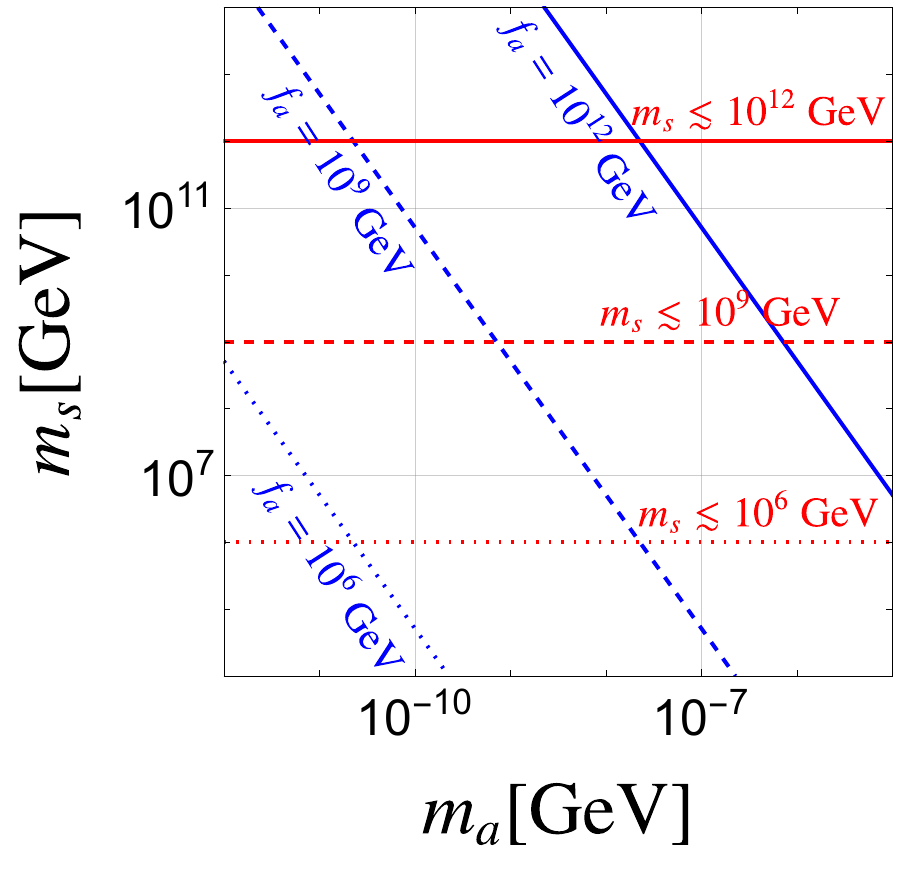}
  \end{tabular}
\end{center}
\caption{The contour plot of the free-streaming length $\lambda_{fs}= 0.1\mpc$ in the plane of the axion DM mass $m_a$ and the saxion mass $m_s$,  for $f_a=({10^{6},10^{9},10^{12}})\gev$ with blue lines  (dotted, dashed, and solid), respectively. The constraint $m_s\lesssim f_a$ is shown with red lines for each $f_a$ given on the line.  We can find the minimal mass of the axion DM at the cross point of the blue and red lines for a given $f_a$.}
\label{fig:Lfs_ALP}
\end{figure}
%%%%%%%%%%%%%%%%%%%%%

In Fig.~\ref{fig:Lfs_ALP}, we show the contour plot of $\lfs$ considering that the non-thermal axions make 100\% DM. This can be achieved for appropriate initial density of saxion, as explained in Eq.~(\ref{Oh2_hidden}).  Considering that $m_s\lesssim f_a$, we can find the minimal mass of cold axion DM when this inequality saturates:
\begin{equation}
m_a \gtrsim 10\unitev \, \bfrac{0.1\mpc}{L_\alpha}  \bfrac{f_a}{10^{12}\gev}^{1/2} \, .
\end{equation}
For $f_a=10$ TeV, the axion mass can be as light as 1 meV.

Here we assume that the saxion does not couple to the visible sector and there would be no constraint from the thermalization of the axion DM. The axion DM produced from saxion may cause the problem of the isocurvature perturbation. However, it is possible that if the saxion is very heavy during inflation, its perturbation is exponentially suppressed.

\subsection{DM from $Q$-ball decay}

If the potential of a complex scalar field $\phi$ with a well-conserved global charge is flatter than quadratic one, a non-topological soliton called $Q$-ball may be formed~\cite{Coleman:1985ki}.
If the charge of $Q$-balls is large enough to survive from evaporation, $Q$-ball could decay into lighter particles through Yukawa interactions and/or a scalar trilinear interaction if the decay is kinematically allowed. The $Q$-ball made of $\phi$ with energy $E_{\phi}=\omega$, where $\omega$ is angular velocity of the $Q$-ball solution, decays into two scalar DM particles with the energy $E_{\chi}=\omega/2$. We note that $\omega$ is not necessarily the same as $m_{\phi}$, which is the difference from the coherent oscillation of $\phi$ discussed in the previous example. The background temperature when $Q$-balls decay with the rate $\Gamma_Q$ is estimated as
\begin{align}
T_{\rm decay} = \left(\frac{90}{\pi^2g_*}\right)^{1/4} \mpl^{1/2}\Gamma_Q^{1/2} \, .
\end{align}
If the kinetic energy of $\chi$ could be much smaller than the background radiation for $E_{\chi}=\omega/2 \ll T_{\rm decay}$, then $\chi$ can be LCDM even if it is relativistic at the moment of production. For some given model, the lower bound on the mass of $\chi$ can be found in terms of $\omega$ and $\Gamma_Q$ as
\begin{equation}
\label{eq:Qball-DMmass}
m_\chi \gtrsim \kev \, \bfrac{\omega}{\gev} \bfrac{10^{-18}\gev}{\Gamma_Q}^{1/2} \, .
\end{equation}

Having discussed general idea, now let us see how the parameters for the $Q$-balls are related to $\lfs$. At first, for simplicity, let us approximate the $Q$-ball configuration by the step-function as
\begin{equation}
 \phi(r) = 
\left\{
\begin{array}{cc}
\phi_0  & (r<R)
\\
0  &    (r>R)
\end{array}
\right. 
\, .
\label{Eq:phi(r)}
\end{equation}
With its radius $R$, the volume of the $Q$ ball is given by $V_Q=4\pi R^3/3$. 
The interaction parts of $\phi$ in Lagrangian may be given by
\begin{equation}
\calL_\mathrm{int} = - \psi_1^{\dagger} \big( g\phi \psi_0 +y\varphi_1\psi_0^{\prime} \big)
  - y\varphi_0 \psi_0^T \psi_0^{\prime}+ h.c. \, .
\end{equation}
as a simplified example. Here, the fermions $\psi_1$, $\psi_0$ and $\psi_0^{\prime}$ and the scalars $\varphi_0$ and $\varphi_1$ are light final states and their subscripts denote their global $U(1)$ charge. As $\phi$ develops the vacuum expectation value $\langle\phi\rangle$ inside a $Q$-ball, $\psi$ has a large Dirac mass $g\langle\phi\rangle$. Thus, the $Q$-ball can decay into $\psi$ only at the surface. For a large $\langle\phi\rangle$ so that $g \langle\phi\rangle / \omega \gg 1$, the decay rate into fermions is saturated due to the Pauli exclusion~\cite{Cohen:1986ct} as
\begin{equation}
-\left(\frac{dQ}{dt}\right)_{Q \rightarrow \psi\bar{\psi}}
 = -\left(\frac{dQ}{dt} \right)_{\mathrm{sat}} \equiv   \frac{\omega^3 R^2}{24\pi} \, .
\label{Eq:dNdt}
\end{equation}
The decay rate for a more general profile can be found in Ref.~\cite{Kawasaki:2012gk}. If $g \langle\phi\rangle / \omega \ll 1$, the $Q$-ball decay into fermions is not kinematically suppressed. Hence, the dominant scalar DM production from the $Q$-ball decay we are interested in is difficult via the fermion channel. We note that an example of fermionic warm DM from the $Q$-ball decay can be found in Ref.~\cite{Seto:2007ym}.

Next, we consider the $Q$-ball decay into scalars $\varphi_1$ and $\varphi_0$. The decay rate of $\phi$ in the non-vanishing $\phi$ background is, for $m_{\varphi_1},m_{\varphi_0} \ll \omega$, given by
\begin{align}
& \Gamma(\phi\rightarrow \varphi_1 \varphi_0) 
\nonumber \\
& = \frac{\sqrt{(\omega^2-(m_{\varphi_1}+m_{\varphi_0})^2) \big[\omega^2-(m_{\varphi_1}-m_{\varphi_0})^2 \big]}}{16\pi\omega^3}\left|\frac{g y^2}{16\pi^2} m F\right|^2 
 \, ,
\label{Eq:Gammaphi}
\end{align}
where the auxiliary function $F$, defined in the appendix, in the loop calculation with $m=g\langle\phi\rangle$ asymptotically behaves as
\begin{equation}
F \simeq \frac{5}{18}\frac{\omega^2}{m^2} \, ,
\end{equation}
for $m_{\varphi_1},m_{\varphi_0} \ll \omega \ll m$. Integrating Eq.~\eqref{Eq:Gammaphi} over the $Q$-ball background \eqref{Eq:phi(r)}, we obtain the decay rate of $Q$-ball into light scalar as
\begin{align}
-\left(\frac{dQ}{dt}\right)_{Q \rightarrow \varphi_1 \varphi_0} 
& =
4\pi \int^{R} \omega \phi(r)^2 r^2 dr \Gamma(\phi\rightarrow \varphi_1 \varphi_0) 
\nonumber\\
& = Q \Gamma(\phi\rightarrow \varphi_1 \varphi_0) \, .
\label{Eq:dQdt}
\end{align}
The ratio of the $Q$-ball decay into scalars to that into fermions is
\begin{equation}
\frac{\big(dQ/dt\big)_{Q\rightarrow \varphi_1 \varphi_0}}{\big(dQ/dt\big)_{Q \rightarrow \psi\bar{\psi}} } 
= \frac{32\pi^2 \phi_0^2 R}{\omega^2} \Gamma(\phi\rightarrow \varphi_1 \varphi_0) 
= \mathcal{O}(\omega R) \, .
\end{equation}
Thus the decay is dominantly into scalars for a large $Q$-ball with the radius $R \gg 1/\omega$. From Eq.~\eqref{Eq:dQdt}, we find the $Q$-ball decay rate $\Gamma_{Q}$ is just 
\begin{equation} 
\Gamma_{Q} = \Gamma(\phi\rightarrow \varphi_1 \varphi_0) \, .
\end{equation}

For a more realistic configuration than Eq.~\eqref{Eq:phi(r)}, we can adapt the Gaussian profile as 
\begin{equation}
 \phi(r) = \phi_0 e^{-\frac{r^2}{2R^2}} ,
\label{Eq:phi(r)_in_Qball}
\end{equation}
which can be realized by the potential\footnote{
In fact, this potential appears in supergravity model and has been studied extensively (see for e.g.~\cite{Enqvist:2003gh} and references therein). Nevertheless we do not restrict our study for supersymmetric models.}
\begin{equation}
V(\phi) = m_{\phi}^2\left[1+K \log\left(\frac{|\phi|^2}{\Lambda^2}\right)\right]|\phi|^2 + \cdots \, ,
\label{eq:Qball:potential}
\end{equation}
where $K<0$ and $\Lambda$ are some scales, and ellipsis denotes higher order terms not relevant for a $Q$-ball solution but responsible to the global $U(1)$ charge generation. The $Q$-ball radius $R$, $\omega$, the charge of the $Q$-ball $Q$, and $\phi_0$ are given by~\cite{Enqvist:1998en}
\begin{equation}
\begin{split}
R &\simeq \frac{1}{m_{\phi}\sqrt{|K|}} \, , 
\\
\omega &\simeq m_{\phi} \, , 
\\
Q & \simeq \left(\frac{\pi}{2}\right)^{3/2}\omega \phi_0^2 R^3 \, , 
\\
\phi_0 &\simeq \frac{|K|^{3/4}}{\sqrt{2\pi^{3/2}}} m_{\phi} Q^{1/2} \, .
\end{split}
\end{equation}

\begin{figure}[thb]
\centering
\includegraphics[width=0.4\textwidth]{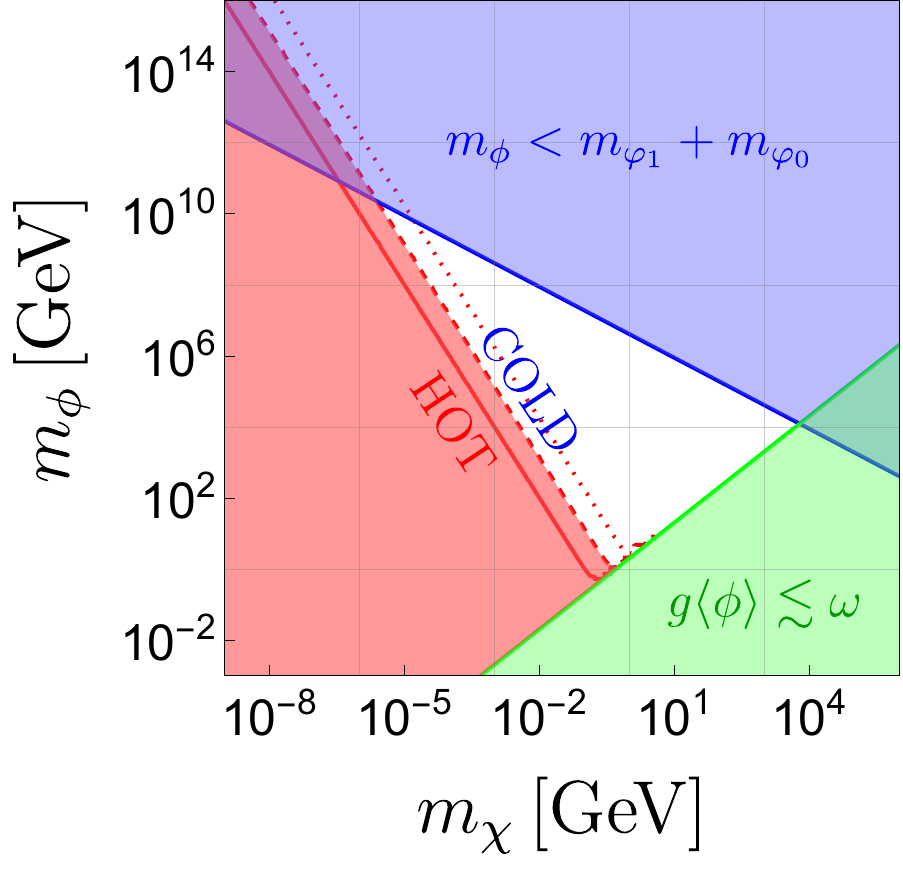}
\caption{
The free-streaming length of DM produced from the $Q$-ball decay.  
The mass spectrum  $m_{\phi} < m_{\varphi_1}+m_{\varphi_0}$ is realized in the top region where the $Q$-ball decay into scalars is not kinematically possible.
The green region is excluded for $g \langle\phi\rangle / \omega \ll 1$ in the lower right corner.
We have used the following values of parameters: $|K|=10^{-4}$, $g=1$, $y=1$ and $m_{CP}=m_{\phi}$. 
 }
\label{Fig:Qballplot}
\end{figure}

Now we estimate the DM abundance, provided that the potential of $\phi$ takes the form of \eqref{eq:Qball:potential}. After inflation, $\phi$ starts to oscillate with an initial amplitude $\phi_i$ when $H^2\simeq m_{\phi}^2$. Then, non-vanishing global $U(1)$ charge of $\phi$ could be generated and the coherent $\phi$ collapses into $Q$-balls. The generated global $U(1)$ asymmetry and the $Q$-ball charge is parameterized respectively as
\begin{align}
n|_{H=m_{\phi}} & \simeq m_{CP} \phi_i^2 \, , 
\label{eq:phiassym} 
\\
Q & \simeq \frac{4\pi}{3}\left(\frac{m_{\phi}}{|K|}\right)^{-3} m_{CP} \phi_i^2 \, ,
\label{eq:Qball:charge}
\end{align}
where $m_{CP}$ is a possible $CP$-violating parameter to generate the global $U(1)$ charge asymmetry. This energy density of $\phi$ converts into that of $Q$-balls, which decay into $\varphi_1$ and $\varphi_0$. The resulting DM abundance is given by
\begin{align}
\frac{\rho_{\chi}}{s} 
& \simeq \left.  \left(\frac{m_{\chi}}{m_{\phi}}\right)\frac{45 m_{\phi}m_{CP}\phi_i^2}{2\pi^2 g_{*S}T^3}\right|_{H=m_{\phi}}
\nonumber\\ 
& \simeq \left(\frac{m_{\chi}}{m_{\phi}}\right)\frac{45 }{2\pi^2 g_{*S}}\left(\frac{90}{\pi^2 g_*}\right)^{-3/4}
 \frac{ m_{CP}\phi_i^2}{m_{\phi}^{1/2}\mpl^{3/2} } \, ,
\end{align}
with $\chi$ denoting $\varphi_1+\varphi_0$ and $m_\chi=m_{\varphi_1}=m_{\varphi_0}$. We demand $\Omega_\chi h^2 = (\rho_{\chi}/s)/(\rho_{\mathrm{cri}}/s) \simeq 0.12$ to meet the observed DM abundance, with $\rho_{\mathrm{cri}}$ being the critical density.
An available parameter region for $|K|=10^{-4}$, $g=1$, $y=1$ and $m_{CP}=m_{\phi}$ is shown in Fig.~\ref{Fig:Qballplot}. Then, we find $m_{\chi} \gtrsim 10$ keV. We note that the lower bound of $m_{\chi}$ increases as $|K|$ becomes larger.

%%%%%%%%%%%%%%%%%%%   
%%%%%%%%%%%%%%%%%%%%%

%%%%%%%%%%%%%%
\section{Conclusion}
\label{sec:con}

Thermally produced particles can be CDM with mass larger than around keV. However, non-thermally produced particles can be cold even with lighter masses. In this article, we have examined systematically such ``light'' CDM, or simply LCDM, produced from the decay of heavy non-thermal particles.

When LCDM is produced from the decay of heavy particle which dominates the energy of the Universe when it decays, e.g. the inflaton, the lowest possible mass of LCDM is around $10 \unitev$. However, when the heavy particle decays dominantly to hidden sector, the mass of LCDM can be much lighter.

As specific examples, we have considered the axion DM from heavy saxion decay and scalar DM from the $Q$-ball decay. For axion DM, the mass can be small down to $10^{-3}$eV for the PQ scale around TeV. However, DM produced from the $Q$-ball can be light around keV.

%%%%%%%%%%%%%%

\section*{Acknowlegments}

This work is supported by the National Research Foundation of Korea grants 2022R1A2C1005050 (KYC, JJ), 2019R1A 2C2085023 (JG), 2017R1D1A1B06035959 (WIP) and 2022R1 A4A5030362 (WIP). JG is also supported by the Ewha Womans University Research Grant of 2022 (1-2022-0606-001-1). WIP is also supported by the Spanish grants PID2020-113775GB-I00 (AEI/10.13039/501100011033) and CIPROM/2021/054 (Generalitat Valenciana).
OS is supported in part by the Japan Society for the Promotion of Science KAKENHI Grants No.~19K03860, No.~19K03865 and No.~23K03402.
We also acknowledge the Korea-Japan Bilateral Open Partnership Joint Research Projects supported by the National Research Foundation of Korea and the Japan Society for the Promotion of Science (2020K2A9A2A 08000097).
JG is grateful to the Asia Pacific Center for Theoretical Physics for hospitality while this work was under progress.

\appendix

\section{Auxiliary function}

The amplitude for the decay $\phi(p) \rightarrow \varphi_1(q) \varphi_0(p-q)$ is expressed as
\begin{equation}
i\mathcal{M}  = \frac{g y^2}{16\pi^2} F \, , 
\end{equation}
where the auxiliary function is defined by
\begin{equation}
F \equiv \int_0^1\int_0^1 2ydydx \frac{2l^2+(p+2q)\cdot l+p\cdot q}{l^2 +q^2(-1+y)+\big[m^2+p^2(-1+x)\big]y} \, , 
\end{equation}
and, for $m_{\varphi_1},m_{\varphi_0} \ll \omega$, asymptotically 
\begin{align}
F \simeq \frac{5}{18}\frac{\omega^2}{m^2} \, ,
\end{align}
with 
\begin{align}
l^{\mu} &= (-1+y)q^{\mu}+(-1+x)yp^{\mu} \, , 
\\
m & =g\phi \, .
\end{align}

\bibliographystyle{unsrt}
\bibliography{reference}

\end{document}